# Seismic Interevent Time: A Spatial Scaling and Multifractality

G. Molchan[1,2] and T. Kronrod[1]

[1]*International Institute of Earthquake Prediction Theory and Mathematical Geophysics, Russian Academy of Sciences, Moscow.*
[2]*The Abdus Salam International Center for Theoretical Physics, SAND Group, Trieste, Italy.*
*E-mails:* molchan@mitp.ru (GM); kronrod@mitp.ru (TK).

The optimal scaling problem for the time $t(L \times L)$ between two successive events in a seismogenic cell of size $L$ is considered. The quantity $t(L \times L)$ is defined for a random cell of a grid covering a seismic region $G$. We solve that problem in terms of a multifractal characteristic of epicenters in $G$ known as the tau-function or generalized fractal dimensions; the solution depends on the type of cell randomization. Our theoretical deductions are corroborated by California seismicity with magnitude $M \geq 2$. In other words, the population of waiting time distributions for $L$ = 10-100 km provides positive information on the multifractal nature of seismicity, which impedes the population to be converted into a unified law by scaling. This study is a follow-up of our analysis of power/unified laws for seismicity (see PAGEOPH 162 (2005), 1135 and GJI 162 (2005), 899).

## 1. Introduction

Bak et al. (2002) introduced into seismicity statistics the notion of a unified scaling law as a generalization of the power law. To clarify this term, let us consider a seismic zone $G$ covered by a regular grid of mesh size $L$. Let $\xi(L \times L)$ be a statistic based on seismic events of magnitude $M > m_c$ in an $L \times L$ cell during time $\Delta T$. The statistic $\xi$ generates a unified law, if there is a normalizing constant $a_L$ such that the distribution of $a_L \xi(L \times L)$ when averaged over seismogenic cells is independent of $L$, $m_c$ and $\Delta T$. The type of the averaging has to be specified.

The example considered by Bak et al. (2002) is concerned with the time between successive events, $t(L \times L)$. The distributions were averaged in proportion to the number of events in the $L \times L$ cell, and the normalizing constant $a_L$ had the form $a_L = \lambda_L$, where

$$\lambda_L = c\,10^{-bm_c}(L/L_0)^{d_f}. \qquad (1)$$

Here, $b$ is the $b$-value in the Gutenberg-Richter law, $L_0$ is the external scale, and $d_f$ was treated as a fractal dimension.

Corral (2003) proposed another example in which $\xi_L$ is the rate, $\lambda(L \times L)$, of $M > m_c$ events in an $L \times L$ cell. The distribution of $\lambda(L \times L)/\lambda_L$ over seismogenic cells is treated as a unified law. The general approach relates an $L \times L$ cell to a distribution having a unit jump at the point $\lambda(L \times L) > 0$, these distributions being averaged with equal weights.

The relation between the normalizations in these two examples follows from the equality

$$E\,t(L \times L) = 1/\lambda(L \times L). \qquad (2)$$

Other examples can be found in Baiesi and Paczuski (2004), Lise et al. (2004), Davidsen et al (2005), Davidsen and Paczuski (2005). All such examples are of great interest for a better understanding of seismicity, unless they are corollaries from facts that are already known: the Gutenberg-Richter power law for energy, the Omori attenuation power law in time, and the fractality of events in space.

As a rule, unified laws are approximate and are in need of restrictions on $L$, $m_c$ and $\Delta T$, since the theoretical result by Molchan (2005) asserts that the statistic $t(L \times L)$ generates a unified law at the scales $L < L_0$ only if the rate of $M > m_c$ events is uniform on its support, i.e. $d_f = 0, 1$ or $2$. Real observable seismicity is organized differently. The commonly accepted hypothesis assumes it to be fractal, i.e., $d_f$ to be non-integer. Researchers are not as unanimous concerning the assumption that the spatial distribution of events is a multifractal (see, e.g., pro: Geilikman et al., 1990; Goltz, 1997; Sornette and Ouillon, 2005 and contra: Eneva, 1996; Gonzato et al., 1998). The scepticism stems from the difficulties inherent in estimating the singularity spectrum of a multifractal and in the undoubted difference between a sophisticated abstract notion of multifractality and observed reality. The utmost to be derived from observable data is a statement like this: seismicity looks as a multifractal in a range of scale $\Delta L$. We have inferred this for $M \geq 2$ California events in the range $\Delta L$=10–100 km; a similar inference for $M \geq 3$ seems questionable (Molchan and Kronrod, 2005).

Given that epicenters are multifractal, the choice of the exponent $d_f$ in (1) becomes non unique. From the very representation $\lambda(L \times L) \propto L^d$ it follows that, when $L$ is small, the parameter $d$ plays the part of an indicator of singularity/smoothness for seismicity rate. It is only for a monofractal that the indicator is unique. For this reason the property of multifractality and the existence of unified laws are largely incompatible. However, treating the laws as approximate ones, one can look for the optimal form of scaling for the statistic $\xi(L \times L)$ that aspires to be the unified law. This problem has been solved for $\xi = \lambda(L \times L)$ by Molchan and Kronrod (2005). We use the multifractal formalism to find theoretically the index $d_f$ in (1) for small $L$. It is a function of the weights involved in the contributions of different $L \times L$ cells. (In the examples cited above, the weight of an $L \times L$ cell is proportional to $\lambda^p(L \times L)$ with $p$=1 and $p$=0). The data observed in California ($M$>2 events) have corroborated the predicted values of $d_f$. Thereby we have obtained an independent confirmation for the multifractality of $M \geq 2$ epicenters.

Below we propose a general approach to the optimal spatial scaling of $t(L \times L)$ and $\lambda(L \times L)$ based on the multifractal formalism. In spite of relation (2) connecting them, the two problems of scaling $t(L \times L)$ and $\lambda(L \times L)$ are different. It is sufficient to remark that a simple averaging of $\lambda(L \times L)$ and $E\,t(L \times L) = \lambda^{-1}(L \times L)$ over the cells responds to small values of $\lambda$ differently. In the generic situation therefore, the averages of $\lambda(L \times L)$ and of $t(L \times L)$ are scaled over $L$ differently. Similarly to the case $\xi = \lambda(L \times L)$, we are going to corroborate the efficiency of the multifractal ideology for dealing with the spatial scaling of $t(L \times L)$.

The subsequent text is organized as follows: Section 2 provides the statement of the problem of scaling $\xi(L \times L)$; Section 3 contains basic information about multifractals; Section 4 gives a theoretical solution as to the choice of the scaling index $d_f$ for $t(L \times L)$ and $\lambda(L \times L)$; in Section 5 the prediction of $d_f$ for $t(L \times L)$ is checked using California events with $M \geq 2$. Lastly, Section 6 discusses the nature of the asymptotic distribution of $t(L \times L)$ for small values.

## 2. Statement of the Problem

Suppose $a_L \xi(L \times L)$ is a suitable normalization of the statistic $\xi(L \times L) \geq 0$ such that the averaged distributions of $a_L \xi(L \times L)$ are weakly dependent on the scale parameter $L$. If the unified law is approximately valid, then the averaged distributions of $a_L \xi(L \times L)$, $F_L$, are close to one another for different $L$, though not exactly coincident. One asks how one is to choose $a_L$ or, as in the case of $t(L \times L)$ and $\lambda(L \times L)$, how one is to choose $d_f(1)$.

It can be assumed theoretically that we have to deal with an infinite set of distributions $\{F_L\}$ in the limit $L \to 0$. One knows (Feller, 1966) that, given any infinite sequence of distributions, one can always select there a subsequence that would converge to a non-decreasing function. The limit may be a constant, corresponding to a distribution that is concentrated at 0 and ∞. A limit of this sort is of no interest whatever. For this reason the normalizing constant $a_L$ or the scaling index $d_f$ should be rejected, if any possible limit for the family $\{F_L(x)\}$ can be a constant only in $0 < x < \infty$. Indices $d_f$ that do not possess the property indicated will be called *suitable*.

Assuming the rate of events to be multifractal, we are faced with the problem of describing all suitable values of $d_f$. If the index is a single one, one can expect $F_L$, $L \to 0$ to converge to a non-trivial limit, hence expect $F_L$ to be close to one another for small $L$.

The definition of a unified law involves the averaging operation which requires to be specified. We shall consider below the following one-parameter family of weights for $L \times L$ cells with $\lambda(L \times L) > 0$:
$$w^{(p)}(L \times L) = [\lambda(L \times L)/\lambda(G)]^p / R_L(p) \qquad (3)$$
where $R_L(p)$ normalizes the weights to unity, and $p$ is a parameter, $|p| < \infty$.

The weights $\{w^{(p)}\}$ can be treated as a probability distribution on earthquake-generating $L \times L$ cells that cover $G$. Let us select at random an $L \times L$ cell using the probability distribution (3) and set $\xi_L^{(p)} = \xi(L \times L)$. Then the random variable $\xi_L^{(p)}$ has the distribution
$$F_L^{(p)}(x) = \sum w^{(p)}(L \times L) F(x \mid L \times L)$$
where $F(x \mid L \times L)$ is the distribution of $\xi(L \times L)$ for the $L \times L$ cell. In other words, the distribution of $\xi_L^{(p)}$ corresponds to the desired averaging for the distributions of $\xi(L \times L)$ with weights (3). When $p=1$ and $\xi = t(L \times L)$, the variable $\xi_L^{(p)}$ has the distribution studied by Bak et al. (2002); the case $p=0$ with $\xi = \lambda(L \times L)$ corresponds to the Corral (2003) example. When $p \gg 1$, $F_L^{(p)}(x)$ is identical with the distribution of $\xi(L \times L)$ in the cell having the greatest rate $\lambda(L \times L)$.

Below we describe suitable values of $d_f$ for the family of weights (3) at any fixed $p \geq 0$. To do this, we shall need some basic facts related to multifractals (see, e.g., Feder, 1988).

## 3. Multifractal Seismicity.

Let the measure $\lambda(dg)$ be the rate of $M > m$ events in an element of area $dg$ in region $G$. The measure $\lambda(dg)$ is treated as a multifractal, if its support stratifies into a set of fractal subsets $S_\alpha$ having Hausdorff dimensions $f(\alpha)$. In addition, the points in $S_\alpha$ possess the following property: for any point $g \in S_\alpha$ there exists a sequence of $L \times L$ cells as $L \to 0$ such that
$$\log \lambda(L \times L) = \alpha \log L (1 + o(1)). \qquad (4)$$

In other words, $\alpha$ describes a type of spatial concentration of events or the singularity type for $\lambda(dg)$. The pairs $(\alpha, f(\alpha))$ form the multifractal spectrum of the measure. Information on the spectrum can be gathered from the Renyi function
$$R_L(q) = \sum \left[\frac{\lambda(L \times L)}{\lambda(G)}\right]^q, \qquad (5)$$
the summation being over all $L \times L$ cells with $\lambda(L \times L) > 0$.

The following asymptotics holds for multifractals:
$$\log R_L(q) = \tau(q) \log L (1 + o(1)), \quad L \to 0 \qquad (6)$$
where the scaling exponent $\tau(q)$ is related to $f(\alpha)$ through the Legendre transformation:
$$\tau(q) = \min_\alpha (q\alpha - f(\alpha)). \qquad (7)$$

The function $\tau(q)$ is concave (see an example in Fig.1) and $\tau(1)=0$, whence $\min(\alpha - f(\alpha))=0$, i.e., $\alpha \geq f(\alpha)$. If $\tau(q)$ is strictly concave and smooth, then the range of values of $\dot{\tau}(q)$ defines the range of possible singularities of the measure, while the Legendre transform
$$f(\alpha) = \min_q (q\alpha - \tau(q))$$
allows the spectrum of the measure to be described through the moment exponents $\tau(q)$. The above statements constitute the content of multifractal formalism which has been justified for a large class of measures interesting for applications (see, e.g., Pesin, 1997).

The quantities $d_q = \tau(q)/(q-1)$ are known as *generalized dimensions*, in particular, $d_0$ is the *box* dimension, $d_1 = \dot{\tau}(1)$ the *information* dimension, and $d_2 = \tau(2)$ the *correlation* dimension. Since $\tau(q)$ is a concave function, $d_q$ does not increase with increasing $q$. One has from $\tau(1)=0$:
$$d_q = \frac{\tau(q) - \tau(1)}{q - 1} = \dot{\tau}(q^*)$$
where $q^*$ lies between 1 and $q$, i.e., $d_q$ and $\dot{\tau}(q)$ differently parameterize singularities of $\lambda(dg)$. By the theorem of Young (1981), if $\lambda(S_\alpha) > 0$, then $\alpha = f(\alpha)$, i.e., the type of singularity also specifies the dimension of points of that type. All solutions of $\alpha = f(\alpha)$ belong to the interval $[\dot{\tau}(1+0), \dot{\tau}(1-0)] = \Delta\alpha$ where $\dot{\tau}(q \pm 0)$ denotes the right (+) and the left (−) derivatives of the concave function $\tau$ at the point $q$. In the regular case, closure of the sum of sets $S_\alpha$, $\alpha \in \Delta\alpha$ defines a closed support of $\lambda(dg)$ and its Hausdorff dimension coincides with the box dimension $d_0$. The support of $\lambda(dg)$ is thus related to the box dimension $d_0 = -\tau(0)$ and the information dimensions defined by the interval
$$[\dot{\tau}(1+0), \dot{\tau}(1-0)] = [d_{1+0}, d_{1-0}].$$

From (4) it follows that in the scaling relation $\lambda(L \times L) \propto L^d$ the parameter $d$ must play the part of a "suitable" singularity for the measure $\lambda(dg)$. The parameter may not be identical with the dimension of the seismicity support, and the choice of it may depend on the problem under consideration, since a multifractal is described by the spectrum of singularities.

## 4. Scaling of $t(L \times L)$: A Theoretical Approach.

The scaling of the random variable $t(L \times L)$ proposed by Bak et al. (2002) may be interpreted as follows. For stationary seismicity one has the relation $E t(L \times L) = 1/\lambda(L \times L)$. Consequently, it is sufficient to scale the rate $\lambda(L \times L)$ for a *typical* $L \times L$ cell. Using the Gutenberg-Richter law and the fractality of seismicity, we arrive at the normalizing function (1). The notion of a typical cell is determined by the choice of the weights $\{w^{(p)}(L \times L)\}$; $p=1$ in the Bak case. The rate $\lambda(L \times L)$ when averaged using the weights $w^{(1)}(L \times L)$ is scaled with the index $d_f = d_2$ (see below), which is the choice of Bak et al. (2002). These arguments are rather crude for multifractal seismicity. We are going to show that the scaling of the means of $\lambda(L \times L)$ and of $t(L \times L)$ are different, and are also different from the scaling of the distributions of these variables.

**Scaling of the means**. The sampling of an $L \times L$ cell will be based on the use of the weights $w^{(p)}(L \times L)$ as given by (3). The mean thus weighted will be denoted $\langle \cdot \rangle_p$; $t_L^{(p)} = t(L \times L)$ with probability $w^{(p)}(L \times L)$.

Let us find the mean $E t_L^{(p)}$. One has
$$E t_L^{(p)} = \langle E t(L \times L) \rangle_p = \langle \lambda^{-1}(L \times L) \rangle_p$$

For an arbitrary $\varepsilon$
$$\langle \lambda^{\varepsilon}(L \times L) \rangle_p = \lambda^{\varepsilon}(G) \sum \left[ \frac{\lambda(L \times L)}{\lambda(G)} \right]^{p+\varepsilon} / R_L(p)$$
$$= \lambda^{\varepsilon}(G) R_L(p+\varepsilon) / R_L(p) \quad .$$

Using (6), one has
$$\log \langle \lambda^{\varepsilon}(L \times L) \rangle_p = [\tau(p) - \tau(p+\varepsilon)] \log L^{-1}(1 + o(1)), \quad L \to 0.$$

Setting $\varepsilon = -1$, one has $E t_L^{(p)} \sim L^{-d_t^{(p)}}$ where
$$d_t^{(p)} = \tau(p) - \tau(p-1). \tag{8}$$

Setting $\varepsilon = 1$, one gets $\langle \lambda(L \times L) \rangle_p \sim L^{d_\lambda^{(p)}}$ where
$$d_\lambda^{(p)} = \tau(p+1) - \tau(p). \tag{9}$$

Using the concavity of $\tau(p)$, one has
$$d_\lambda^{(p)} \leq \dot\tau(p+0) \leq \dot\tau(p-0) \leq d_t^{(p)}. \tag{10}$$

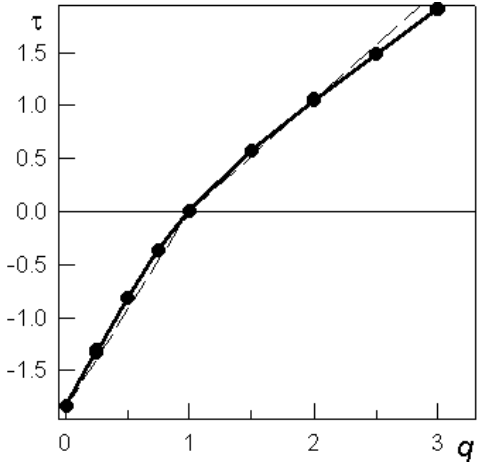

Figure 1. Tau-function for $M \geq 2$ California events (*solid line*) and its bi-fractal approximation (*broken line*); $\tau(p)$ is based on the range of scales $L = 10$–$100$ km.

Molchan and Kronrod (2005) estimated the function $\tau(p)$, $0 \leq p \leq 3$ and its derivatives from $M \geq 2$ California seismicity for the scale range $10$–$100$ km (Fig. 1). This allows more specific forms of (8–10) for the practically important values of $p$:

Table 1

| $p$ | 0 | 1 | 2 |
|---|---|---|---|
| $d_t^{(p)}$ | $\geq 2$ *) | $1.8 \pm 0.1$ | $1.1 \pm 0.05$ |
| $d_\lambda^{(p)}$ | $1.8 \pm 0.1$ | $1.1 \pm 0.05$ | $0.8 \pm 0.05$ |
| $\dot\tau(p)$ | $2 \pm 0.1$ | $1.35 \pm 0.05$ | $0.9 \pm 0.05$ |
| $\tau(p)$ | $-1.8 \pm 0.1$ | $0$ | $1.1 \pm 0.05$ |

*) we use here the inequality $d_\tau^{(0)} \geq \dot\tau(0) \cong 2$, because it is difficult to estimate $\tau(-1)$.

***Scaling of the distributions.*** We now describe the indices for which the distributions
$$\sum w^{(p)}(L \times L) P(t(L \times L) L^d < x) = F_L^{(p)}(xL^{-d}) \tag{11}$$
can have limits at $L=0$ that are different from constants. We will show that *the suitable indices $d_f$ in (11) belong to the interval*
$$\dot\tau(p+0) \leq d_f \leq \dot\tau(p-0), \tag{12}$$
*i.e., $d_f$ is defined uniquely, if $\dot\tau(p)$ is continuous.*

Our considerations are heuristic in character, because a rigorous approach is only possible for a formalized seismicity model. Our inferences can be checked by real seismicity.

Consider the weights $w^{(p)}(L \times L)$ for cells of type $\alpha$ where $\lambda(L \times L) \sim L^{\alpha}$. The use of (6) gives
$$w^{(p)}(L \times L) = \left[ \frac{\lambda(L \times L)}{\lambda(G)} \right]^p / R_L(p) \sim L^{\alpha p - \tau(p)}, \quad L \to 0.$$

We have denoted by $f(\alpha)$ the singularity spectrum of $\lambda(dg)$; consequently, the number of type $\alpha$ cells is increasing like $L^{-f(\alpha)}$. Hence the contribution of type $\alpha$ cells into (11) has the form
$$c_{\alpha,L} L^{\alpha p - f(\alpha) - \tau(p)} F_{\alpha,L}(xL^{-d})$$
where $F_{\alpha,L}(t)$ is the distribution of $t(L \times L)$ when averaged with equal weights over all type $\alpha$ cells, and $\log c_{\alpha,L} = O(1)$ as $L \to 0$.

From (7) it follows that $\alpha p - f(\alpha) - \tau(p) \geq 0$. The equality is attained for the value of $\alpha$ at which $\alpha p - f(\alpha)$ reaches its minimum, i.e.,
$$\dot\tau(p+0) \leq \alpha \leq \dot\tau(p-0) \tag{13}$$
(see Section 3). The sum (11) is thus reduced to the sum $\sum_\alpha c_{\alpha,L} F_{\alpha,L}(xL^{-d})$ as $L \to 0$, where $\alpha$ belongs to the interval (13).

Consider $\alpha \in [\dot\tau(p+0), \dot\tau(p-0)]$. For a type $\alpha$ cell one has
$$t(L \times L) = O(\lambda^{-1}(L \times L)) = O(L^{-\alpha}), \quad L \to 0$$
Consequently, the quantity $L^d t(L \times L)$ converges either to 0 or to $\infty$, if $d > \alpha$ or $d < \alpha$, respectively. It follows that the equality $d = \alpha$ is a necessary condition for the possible limits of $F_{\alpha,L}(xL^{-d})$ not to be constants. In that case, however, $d$ must belong to (13), which coincides with the desired statement (12).

The statement about the equality $d = \alpha$ can be made rigorous by using the Chebyshev inequality (Feller, 1968) and the a priori estimates $L^{\alpha+\delta} < \lambda(L \times L) < L^{\alpha-\delta}$ and $E[t(L \times L) \lambda(L \times L)]^{-\varepsilon} < AL^{-\rho}$, where $\varepsilon, \delta, \rho$ are small numbers and $L \ll 1$.

***Remarks:***
- It follows from the above reasoning in favor of (12) that the weights $w^{(p)}$ in the limit $L \to 0$ act as a filter on the sets $S_\alpha$. The filtering yields the single set $S_\alpha$, $\alpha = \dot\tau(p)$, provided the derivative $\dot\tau(p)$ is defined at $p$. It explains the mechanism of normalization applied to the distribution of $t_L^{(p)}$. The filtered-out set $S_\alpha$ is monofractal, hence $t(L \times L) \sim 1/\lambda(L \times L) \sim L^{-\alpha}$ on it. For the same reason the interval (12) also specifies admissible values of $d_f$ for scaling $\lambda_L^{(p)}$ (Molchan and Kronrod, 2005); $\lambda_L^{(p)}$ equals $\lambda(L \times L)$ with probability $w^{(p)}(L \times L)$.

- The inequalities (10) show that the indices for scaling $t_L^{(p)}$ and $\lambda_L^{(p)}$ lie between those for their means and need not coincide with them. The fact is not obvious beforehand.

***Example.***

We provide a simple example in which the distributions of $t_L^{(p)} L^d$ with $d = \dot\tau(p)$ have a limit as $L \to 0$. That means that the distributions of $t_L^{(p)}$ will be close to one another for small $L$. It will follow from the example that the limiting distributions or their tails are not universal.

Consider a poissonian field of events on the set $G = I_1 \cup I_2$ consisting of an interval $I_1$ and a square $I_2$. Suppose the rate of events $\lambda(dg)$ has finite density $\dot\lambda(g)$ on each part of $G$. Then the measure $\lambda(dg)$ is a simple mixture of monofractals, so that the spectrum $(\alpha, f(\alpha))$ consists of two points, $(1,1)$ and $(2,2)$.

Accordingly, $\tau(p)$, $\tau(1) = 0$, is a piecewise linear function with $\dot\tau(p) = 2$ for $p<1$ and with $\dot\tau(p) = 1$ for $p>1$. The distribution of $t(L\times L)$ is exponential, i.e.,

$P(t(L\times L)) \lambda(L\times L)) > x) = \exp(-x)$.

Hence the normalized quantity $t_L^{(p)} L^d$ for $p \neq 1$ and $d = \dot\tau(p)$ has the following probability density function (p.d.f.) in the limit $L \to 0$:

$$f^{(p)}(t) = \int_{I_d} \dot\lambda^{p+1}(g) e^{-\dot\lambda(g)t} dg / k = \int x^{p+1} e^{-xt} dF^{(d)}(x)/k \quad (14)$$

where $d = \dot\tau(p)$, $k = \int_{I_d} \dot\lambda^p(g) dg$ is the normalizing constant, and $F^{(d)}$ is the distribution of $\dot\lambda(g)$ on $I_d$.

When $p=1$, two normalizations (with $d=1=\dot\tau(1+0)$ and $d=2=\dot\tau(1-0)$) are possible; the corresponding limits have the form (14) with $k$ modified: $k=\lambda(G)$. When $p=1$, the integral of (14) is less than 1, because a $\delta$-function appears at 0 when $d=2$ and at $\infty$, when $d=1$.

The example does not involve aftershocks, and the p.d.f. of $t(L\times L)$ is bounded around 0. This does not however rule out that the p.d.f. of the limiting distribution may have a singularity at $t=0$ if $\dot\lambda(g)$ is unbounded function. Assuming $1-F^{(d)}(x) \sim cx^\beta$, $x \gg 1$ and $0 < \beta - p < 1$, we find in virtue of the Tauber theorem (Feller, 1966) that

$f^{(p)}(t) \cong c_1 t^{-1+(\beta-p)}$, $t \ll 1$. \quad (15)

Suppose that $\dot\lambda(g) \cong c|g-g_0|^{-d/\beta}$, $\beta>1$, near a point $g_0 \in I_d$, $d=1$ or 2. Then (15) holds and the multifractal spectrum has additional point $(\alpha_0, f(\alpha_0) = 0)$ where $\alpha_0 = d(1-1/\beta)$. Assuming $p=1$ one has $0<\alpha_0<d/2$, $\alpha_0 = \dot\tau(\infty)$ and $tf^{(1)}(t) \propto t^{\alpha_0/(d-\alpha_0)}$. Note that $f^{(p)}(t)=c\exp(-ct)$ and $\tau(0) = \alpha_0 p$ for $p>(1-\alpha_0)^{-1}$.

### 5. Scaling of $t(L\times L)$: An Empirical Approach.

We used the ANSS (2004) catalog of $M\geq 2$ California events within 100 km depth for the period 1984-2004. Two events were counted as one, when the spatial and time distances between the two did not exceed 1 km and 40 seconds, respectively. The spatial region $G$ can be seen in Fig. 2; its linear size is $L_0 = \sqrt{areaG} = 998.16$ km, and the total number of events in $G$ is $N=125144$.

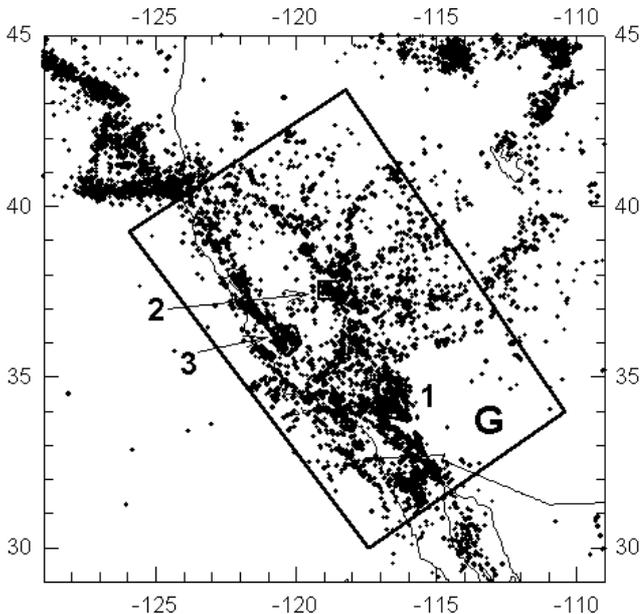

Figure 2. Map of $M\geq 3$ California events and the test area $G$ (rectangle). The polygons 1–3 correspond to most intensive aftershock zones: 1–Landers 1992, $M=7.3$; Big Bear 1992, $M=6.3$; North Palm Springs 1986, $M=6.0$; 2–Chalfant Valley, 1986, $M=6.4$; 3–Coalinga, 1983, $M=6.5$.

The multifractal characteristic of epicenters, the $\tau(p)$-function (Fig. 1), was found by these authors (2005) for $0\leq p\leq 3$ using the scale range $\Delta L=10$–100 km. Values of practical interest for $\tau(p)$ and $\dot\tau(p)$ are listed in Table 1.

Since (6) holds in the range $\Delta L$ and $\tau(p)$, $0\leq p\leq 3$ is nonlinear, we infer that the rate of $M\geq 2$ events in $G$ looks as a multifractal in the scale range 10–100 km (Molchan and Kronrod, 2005).

Below we consider averaging with the weights $w^{(p)}$ and $p=1, 2$; the theoretical values of $d$ are then given by

$\dot\tau(1) = 1.35 \pm 0.05$, $\dot\tau(2) = 0.9 \pm 0.05$ \quad (16)

The variant with equal weights ($p=0$) is not considered because of the great effect of half-empty $L\times L$ cells for which the distributions of $t(L\times L)$ are poorly determined.

The quantity $t(L\times L)$ varies through five orders of magnitude. For this reason it is usually considered on the log-scale, i.e., distributions of $\lg(t_L^{(p)} L^d)$ will be studied. Note that, if $f_\xi$ is the probability density function (p.d.f.) of $\xi$, then the plots of $(x, f_{\lg\xi}(x))$ and $(\lg u, uf_\xi(u))$ are identical.

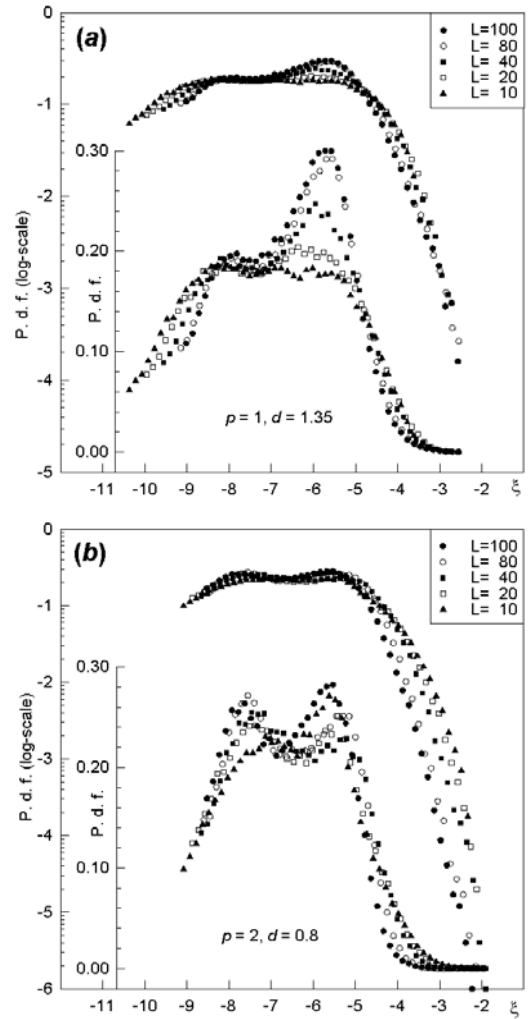

Figure 3. Probability density functions of $\xi = \lg[ct_L^{(p)} \cdot (L/L_0)^d]$ on two scales: logarithmic (upper plots) and usual (lower plots). Parameters: (a) $p=1$, $d=1.35$ and (b) $p=2$, $d=0.8$; $\lg c = -7.8051$.

Figure 3 shows the p.d.f. of $\eta_{p,d} = \lg[t_L^{(p)} L^d c]$ for $(p, d)$ pairs: (a) (1, 1.35) and (b) (2, 0.8), which are close to (16). They are plotted on two vertical scales: the ordinary and the log scale. When seen on the log scale, one usually notes a flat region in the p.d.f. of $\eta_{p,d}$, $p=1$ for small times ($\eta<-6$) and an exponential part for large times (Bak et al., 2002). This is also borne out by Fig. 3. The ordinary scale and different $p$ are not popular when deriving unified laws, although the differences in p.d.f. for different $L$ and $p$ are obvious. We recall that the weight of seismi-city clusters increases with increasing $p$. For moderate values of $L$ ($L>40$ km) the space-time clusters are not yet destroyed by the $L{\times}L$ grid. As a result, the density of $\eta_{p,d}$ for these $L$ is bimodal if $p=2$ and unimodal if $p=1$. This kind of behavior is not affected by selecting $d$ in the interval (0.5, 2). The left mode in Fig. 3b (the case $p=2$) disappears, when the three most intensive aftershock zones have been eliminated (see Fig. 1). In other words, the log-log scale provides a simplified representation of how waiting times are distributed.

We note (as being important for what follows) a correlation between empirical distributions of $\eta_{p,d}$ for different $L$. Denote by $\Delta t(L)$ the waiting time after an event $A$ in an $L{\times}L$ cell, and suppose that cell to be embedded in $L_1{\times}L_1$, $L<L_1$. One has $\Delta t(L_1) \leq \Delta t(L)$, because $L_1{\times}L_1$ contains more events. For this reason the quantities

$$t_-(L) = \min_{L\times L} t(L{\times}L) \quad \text{and} \quad t_+(L) = \max_{L\times L} t(L{\times}L)$$

which define the boundaries of the empirical distribution of $t_L^{(p)}$, must decrease with increasing $L$. The situation changes after the normalization of $t(L{\times}L)$, because the factor $L^d$ increases with $L$. The effect of normalization can be better understood by remarking that the minimum values of $t(L{\times}L)$ are controlled by aftershocks, while the maximum ones by areas of diffuse seismicity. For this reason the values of $t_-(L)$ are close to one another for $L=10$–$100$ km, while the $t_+(L)$ are subject to greater scatter, being in the order $t_+(100),\ldots,t_+(10)$. As a result of the normalization, the left boundary $t_-(L)$ will be displaced on the log scale relative to $t_-(L=100)$ by the amount

$$\delta = -d \log(L^*/L), \quad L^* = 100\,\text{km}.$$

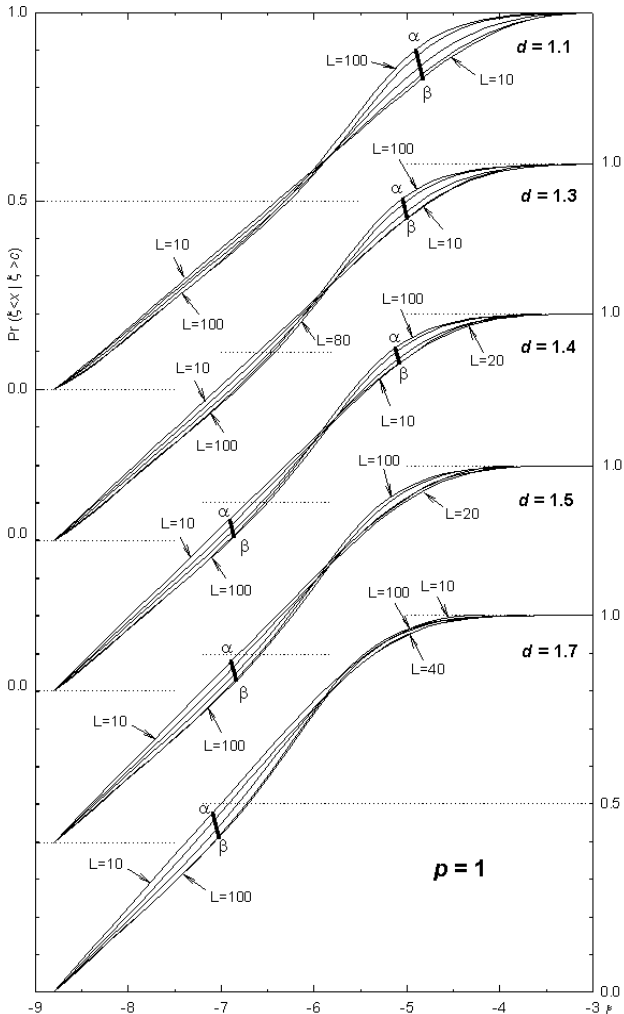

Figure 4. Conditional distribution functions of $\xi = \lg[\,c t_L^{(p)} \cdot (L/L_0)^d\,]$ with $p=1$, $d=1.1$–$1.7$ and $L=10, 20, 40, 80, 100$ km. $(\alpha, \beta)$ is Levy's distance for a set of distributions.

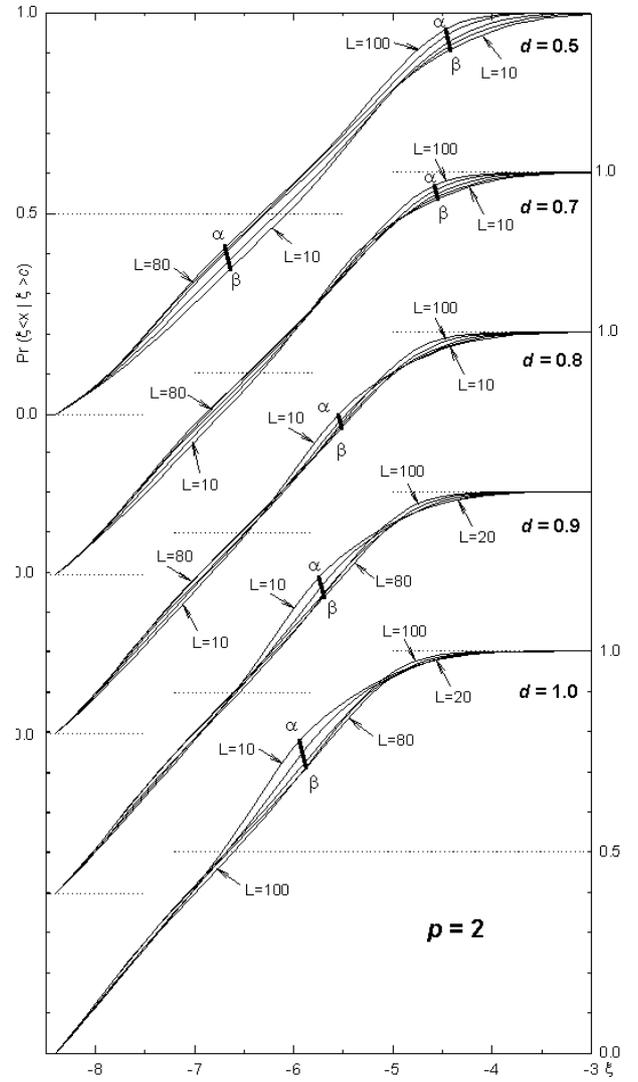

Figure 5. Conditional distribution functions of $\xi = \lg[\,c t_L^{(p)} \cdot (L/L_0)^d\,]$ with $p=2$, $d=0.5$–$1.0$ and $L=10, 20, 40, 80, 100$ km. $(\alpha, \beta)$ is Levy's distance for a set of distributions.

The scatter will be the smaller, the smaller is $d$ (cf. Figs. 3a and 3b). The situation for the right boundary $t_+(L)$ is directly opposite: in order to surmount the large scatter in $t_+(L)$ the relative shift $\delta$ must be substantial, i.e., large $d$ should be used (cf. Figs. 3a and 3b).

The parameter $d$ is thus an important one in statements concerning the universality of the left/right tails in the waiting time distribution (see Davidsen and Goltz, 2004; Corral, 2003).

***The choice of $d$.*** One can see that the empirical distributions of $t_L^{(p)}$ for different $L$ are forced to start from different points. For this reason we shall compare conditional distributions of $t_L^{(p)} L^d$ under the additional condition that $t_L^{(p)} L^d > c$ where $c = \min_{L\times L} t(L{\times}L) L^d$, $L=100$ km. The differences in the right-hand end-points are not so important, because the distributions of $t_L^{(p)}$ decay exponentially for large values.

Empirical conditional distributions of $\log(t_L^{(p)} L^d)$ are shown in Fig. 4 (the case $p=1$) and in Fig. 5 ($p=2$). The distributions are displayed in sequences for $L=10, 20, 40, 80$ and $100$ km and for different $d$. The parameter $d$ was varied around the theoretical values (16): 1.1-1.7 for $p=1$ and 0.5-1.0 for $p=2$.

The Levy metric $\Delta$ (Feller, 1968) is used to measure how close the distributions in a sequence are. To explain the metric, consider plots of distribution functions in a sequence as sets of points on the (x, y)-plane and find the greatest thickness $\Delta$ of the set in the direction of the (-1, 1)-vector. The extreme thickness is marked by the interval $(\alpha, \beta)$ in Figs. 4 and 5. The length of $(\alpha, \beta)$ is Levy's divergence measure.

The conditional distributions of $\log(t_L^{(p)} L^d)$, $L$=10–100 km are the closest for $d$=1.4 in the case $p$=1 and for $d$=0.8 in the case $p$=2. This is fairly well consistent with the theoretical estimates $d=1.35 \pm 0.05$ and $0.9 \pm 0.05$, respectively, as given for the case of small $L$. The agreement between theoretical and empirical estimates of d may be regarded as an independent argument in favor of the multifractality of $M \geq 2$ seismicity in California at the scales $L$=10–100 km.

### 6. The Distribution of t(L×L) for Small Values.

The work of Bak et al. (2002) has drawn attention to the question of how the distribution of waiting time behaves for small values of t (see, e.g., Corral (2003), Davidsen and Goltz (2004), Molchan (2005)). The question has proved to be difficult and requires some discussion.

If an $L \times L$ square has been fixed, then for a wide class of models the probability density of $t(L \times L)$ mimics the Omori law for small $t$ (see the rigorous result by Molchan (2005)). It is doubtful whether that result can be transferred to the statistic $t_L^{(1)}$, since there will be two small parameters then, $t$ and $L$, so that the result may be affected by the order of the limiting processes. To corroborate this, let us consider the case where an $L \times L$ cell is narrowed to become a point $g$. This narrowing of the cell diminishes the flow of events around g. When the events are not strongly correlated over time, the diminishing flow must be nearly poissonian. Examples like this one can be found in the book of Daley and Vere-Jones (2003). This intuitive consideration is proved rigorously in the Appendix for the simplest model which involves main events, aftershocks and the Omori law. The chief inference consists in the relation

$$P(t(L \times L) \lambda(L \times L) > t) \cong \exp(-t), \quad L \ll 1. \quad (17)$$

Some extra requirements are to be imposed in order to make (17) uniform over $t$. In that case we are in the situation considered in the example of Section 4. When (17) holds exactly, we have found the limiting distribution of the normalized $t_L^{(p)}$ and pointed out the conditions under which the p.d.f. of $t_L^{(p)}$ has a power-law singularity at zero. The exponent of this power asymptotic is not universal (see an opposite opinion in (Davidsen and Goltz, 2004)). The statistical difficulties inherent in the analysis of small/large deviations and the potential dependence of the answer on the order of the limiting processes in $t$ and $L$ constitute the main sources of contradictory assertions as to the tails of the distribution of $t_L^{(p)}$.

### 7. Discussion and Conclusion.

The main motive of this study is to try to understand the nontriviality of the information provided by the new scaling laws for seismicity, in particular by the unified law for waiting time $t(L \times L)$ of an event in an $L \times L$ cell suggested by Bak et al (2002).

The prerequisites of the law are nothing out of the ordinary, including as they do the following:

– the relation $E t(L \times L) = 1/\lambda(L \times L)$ which holds for stationary seismicity and the scaling $\lambda(L \times L) \propto L^d$ which is valid for a monofractal;

– limit theorems for classical seismicity models that incorporate main shocks, aftershocks and the Omori law; the theoretical results show that the distribution of $t(L \times L) L^d$ when $w^{(p)}$-averaged has a nontrivial limit as $L \to 0$. This ensures that the normalized distributions of $t_L^{(p)}$ are close to one another for small $L$;

– the methods employed to visualize potentially close distributions, namely, (1) the use of a log-log representation for the density in which attention focuses on the asymptotics of the tails and (2) scaling of $t(L \times L)$ with a suitable parameter $d$. This allows shifting the distributions on the log scale differently so that one of the tails collapses, thus giving an illusion of being independent of $L$.

Also, there are serious obstacles in the way of a unified law for waiting time:

– the law when understood rigorously does not exist for spatially heterogeneous seismicity (Molchan, 2005);

– the multifractality of seismicity is incompatible with the scaling of $t(L \times L)$ by using a single index of singularity $d$;

– the law involves an averaging using weights proportional to $[\lambda(L \times L)]^p$ with $p$=1; the exceptional role of the parameter $p$=1 is by no means clear;

– when the distribution density of waiting time is viewed on the ordinary scale (on the y-axis), one notices the influence of the scales below and above 40 km, as well as the role of clusters of events. The latter circumstance is felt when passing from the weights with $p$=1 to those with $p$=2.

The hierarchical Bak method of analysis of waiting time leads to dismissal of aftershocks (Bak et al, 2002). This important conclusion requires a different method of corroboration due to the difficulties described.

It is more productive to ask about the optimal/admissible spatial scaling of waiting time. We have established a relation between that problem and the multifractal characteristics of seismicity. The spatial scaling index $d_f$ is rigidly related to the type of averaging over the cells (see (12)). Empirical analysis corroborates the theoretical selection of $d_f$. This circumstance is important, since the ideas of multifractality for seismicity have not yet settled down; evaluation of the leading characteristic $\tau(p)$ is stated for unjustifiably long ranges, both of the parameter $p$ and scale $L$, which is detrimental to the confidence one might otherwise have placed in them. A constructive use of $\tau(p)$ to predict the scaling index of waiting times is an independent test of multifractality and argues in its favor.

### *Acknowledgements*
This work was supported by the Russian Foundation for Basic Research (grant 05-05-64384a) and by the European Commission's Project 12975 (NEST) "Extreme Events: Causes and Consequences (E2−C2).

## *Appendix*

### *The model.*

Consider $(t, g)$=(time, location) events. These divide into main shocks and aftershocks. The former constitute a marked Poisson process $x=(t, g, A)$ with the rate measure $m(dx) = dt\, \lambda^*(dg)\, dF(A)$, where $A$ is a function of earthquake magnitude. Each main event $x$ generates a poissonian aftershock cluster $N_0(ds, d\tilde{g}\,|\,x)$ with the conditional rate measure

$$\mu(ds, d\tilde{g}\,|\,x) = Af(s-t)\, ds\, \lambda_0(d\tilde{g}-g), \quad x=(t, g, A).$$

It may be assumed without loss of generality that $\int f(s)\, ds = \int \lambda_0(d\tilde{g}) = 1$ and $EA = \overline{A} < \infty$. The aftershocks for different main events are independent. Seismicity models such as this one date back to Vere-Jones and Davies (1966).

The rate measure of the total flow of events $N(dt, dg)$ is given by the relation

$$\lambda(d\tilde{g})\, dt = \lambda^*(d\tilde{g})\, dt + \lambda^{(a)}(d\tilde{g})\, dt,$$

where

$$\lambda^{(a)}(d\tilde{g}) = \overline{A} \int \lambda_0(d\tilde{g}-g)\, \lambda^*(dg)$$

is the average (over $A$) spatial measure of aftershock rate.

**Statement 1.** *We assume that*

$$1 - F(A) = O(A^{-1-\varepsilon_F}), \quad A \to \infty, \tag{18}$$

$$\max_g \lambda_0(L \times L - g) < c\, L^{\varepsilon_\lambda}, \quad L \to 0, \tag{19}$$

*where $\{L \times L - g\}$ is a shift of the cell by vector $g$. Then*

$$P(t(L \times L)\lambda(L \times L) > t) = \exp\{-t(1 + O(L^\rho))\}(1 + O(L^\rho)), L \to 0,$$

*where*

$$\rho = \varepsilon_F\, \varepsilon_\lambda (1 + \varepsilon_F)^{-1}.$$

*Remark.* If $A$ depends on magnitude as $A=10^{\alpha M}$, then $1-F(A) = cA^{-b/\alpha}$ where $b$ is the b-value in the Gutenberg-Richter law. Usually $b$ is close to 1 and $\alpha$ is between 0.5 and 1 (Sornette and Werner, 2004).

*Proof.*

For a stationary flow of events $N(dt, dg)$ the distribution of $t(L \times L)$ is defined by

$$P(t(L \times L) > t) = -\frac{d}{dt} P\{N(B) = 0\} / \Lambda$$

where $B$ is the set of $(s, g)$ points: $0 < s < t$, $g \in L \times L$, and $\Lambda = \lambda(L \times L)$ (see Daley and Vere-Jones (2003)). The main events constitute a Poisson process, and so

$$P\{N(B) = 0\} = \exp\{-\Lambda^* t - I_1\}$$

where $\Lambda^* = \lambda^*(L \times L)$ and

$$I_1 = \int [1 - \chi_B(x)]\, P\{N_0(B|x) > 0\}\, m(dx),\, x=(s, g, A) \tag{20}$$

Here, $\chi_B = 1$, if $(s, g) \in B$ and 0 otherwise;

$$m(dx) = ds\, \lambda^*(dg)\, dF(A).$$

For a poissonian sequence of aftershocks $N_0(ds, d\tilde{g})$ one has

$$P\{N_0(B\,|\,x) > 0\} = 1 - \exp(-\pi(x)),$$

where

$$\pi(x) = A\, \lambda_0(L \times L - g) \int_0^t f(u - s)\, du.$$

Consequently, for the model under consideration one has

$$P\{t(L \times L) > t\} = \exp\{-\Lambda^* t - I_1\}[\Lambda^* + \dot{I}_1] / \Lambda, \tag{21}$$

where $\dot{I}_1 = \frac{d}{dt} I_1$.

We are going to find the asymptotics for $I_1$ and $\dot{I}_1$ as $\to 0$. By (20),

$$I_1 = \int (1 - e^{-\pi(x)})\, m(dx) - \int \chi_B(x)\, (1 - e^{-\pi(x)})\, m(dx).$$

The second term in $I_1$ can be evaluated as follows:

$$\int \chi_B(x)\, (1 - e^{-\pi(x)})\, m(dx) < \int \chi_B(x)\, \pi(x)\, m(dx)$$

$$= \overline{A} \int_{L \times L} \lambda_0(L \times L - g)\, \lambda^*(dg) \int_0^t ds \int_0^t f(u-s)\, du$$

$$\leq \overline{A} \max_g \lambda_0(L \times L - g)\, \lambda^*(L \times L) \int_0^t \Phi(s)\, ds$$

$$= \lambda^*(L \times L)\, t\, O(L^{\varepsilon_\lambda} \Phi(t))$$

where $\Phi(s) = \int_0^s f(u)\, du \leq \Phi(t)$, $s \leq t$ and the bound (19) has been used. In order to evaluate the first term in $I_1$ we represent it in the form

$$\int (1 - e^{-\pi(x)})\, m(dx) = I + R_1 + R_2 + R_3,$$

where

$$I = \int \pi(x)\, m(dx) = \lambda^{(a)}(L \times L)\, t,$$

$$R_1 = - \int_{A > L^{-\delta}} \pi(x)\, m(dx),$$

$$R_2 = \int_{A < L^{-\delta}} (1 - \pi(x) - e^{-\pi(x)})\, m(dx),$$

$$R_3 = \int_{A > L^{-\delta}} (1 - e^{-\pi(x)})\, m(dx),$$

and $\delta$ is a small parameter that is more conveniently defined later.

Using (18), i.e., the relation $1 - F(A) = O(A^{-1-\varepsilon_F})$, $A \to \infty$, we have

$$|R_1| + R_3 < 2 \int_{A>L^{-\delta}} \pi(x) \, m(dx)$$

$$= 2 \int_{A>L^{-\delta}} A \, dF(A) \int \lambda_0 (L \times L - g) \lambda^*(dg) \int_0^t du \int f(u-s) \, ds$$

$$= O(L^{\delta \varepsilon_F}) \lambda^{(a)}(L \times L) \, t.$$

We now evaluate $R_2$:

$$|R_2| < \frac{1}{2} \int_{A<L^{-\delta}} (\pi(x))^2 \, m(dx)$$

$$= \frac{1}{2} \int_{A<L^{-\delta}} A^2 \, dF(A) \int \lambda_0^2 (L \times L - g) \lambda^*(dg) \int ds \left( \int_0^t f(u-s) \, du \right)^2$$

$$< \frac{1}{2} L^{-\delta} \max_g \lambda_0(L \times L - g) \, \lambda^{(a)}(L \times L) \max_s \int_0^t f(u-s) \, du \cdot t$$

$$= O(L^{\varepsilon - \delta}) \lambda^{(a)}(L \times L) \, t \cdot \max_s (\Phi(t+s) - \Phi(s)).$$

We require $\delta \varepsilon_F = \varepsilon_\lambda - \delta$. Then $\delta = \varepsilon_\lambda / (1 + \varepsilon_F)$. Combining the results for the terms in $I_1$, we get

$$\Lambda^* t + I_1 = \lambda(L \times L) \, t + \lambda^{(a)}(L \times L) \, t \cdot O(L^{\delta \varepsilon_F}) + \lambda^*(L \times L) \, t \cdot O(L^{\varepsilon_\lambda}).$$

Recall that $\lambda^{(a)}(L \times L) / \lambda(L \times L) \leq 1$ and $\lambda^*(L \times L) / \lambda(L \times L) \leq 1$. Hence

$$\Lambda^* t + I_1 = \lambda(L \times L) \, t \, (1 + O(L^{\delta \varepsilon_F})) \tag{22}$$

Here we have used the relation $\delta \varepsilon_F = \varepsilon_\lambda - \delta < \varepsilon_\lambda$.
We now evaluate $\dot{I}_1$. We have

$$\dot{I}_1 = \lambda^{(a)}(L \times L) - R_4 - R_5$$

where

$$R_4 = \int \left[ 1 - e^{-\pi(x)} \right] A \, \lambda_0(L \times L - g) f(t-s) \, m(dx),$$

$$R_5 = \int \chi_B \, e^{-\pi(x)} A \, \lambda_0(L \times L - g) f(t-s) \, m(dx),$$

To evaluate $R_4$ we shall use the inequalities

$$1 - e^{-\pi(x)} < \begin{cases} \pi(x) & \text{if } A < L^{-\delta} \\ 1 & \text{if } A > L^{-\delta}. \end{cases}$$

In that case, in the interval $A < L^{-\delta}$, the quantity $R_4$ is bounded from above by

$$\int_{A<L^{-\delta}} A^2 \, dF(a) \int \lambda_0^2 (L \times L - g) \, \lambda^*(dg) \int ds \, f(t-s) \int_0^t f(u-s) \, du$$

$$< L^{-\delta} \max_g (L \times L - g) \, \lambda^{(a)}(L \times L) \max_s \int_0^t f(u-s) \, du$$

$$= \lambda^{(a)}(L \times L) \cdot O(L^{\varepsilon_\lambda - \delta}).$$

In the interval $A > L^{-\delta}$ the quantity $R_4$ does not exceed

$$\int_{A>L^{-\delta}} A \, dF(a) \cdot \lambda^{(a)}(L \times L) / \overline{A} = O(L^{\delta \varepsilon_F}) \, \lambda^{(a)}(L \times L).$$

To sum up,

$$R_4 = O(L^{\delta \varepsilon_F}) \, \lambda^{(a)}(L \times L).$$

Similarly,

$$R_5 = O(L^{\varepsilon_\lambda}) \, \lambda^*(L \times L) \, \Phi(t).$$

Proceeding as in the derivation of (22), we conclude that

$$(\Lambda^* + \dot{I}_1) / \Lambda = 1 + O(L^{\delta \varepsilon_F}). \tag{23}$$

Relations (21), (22) and (23) prove Statement 1.